\documentstyle[mprocl]{article}

\def\Journal#1#2#3#4{{#1} {\bf #2}, #3 (#4)}

\def\PRL{Phys. Rev. Lett.}
\def\PRD{{Phys. Rev.} D}

\def\be{\begin{equation}}
\def\ee{\end{equation}}
\def\bea{\begin{eqnarray}}
\def\eea{\end{eqnarray}}
\begin{document}

\title{BLACK HOLE THERMODYNAMICS TODAY\footnote{Report of parallel
session chair, to appear in {\it Proceedings of the Eighth Marcel
Grossmann Meeting} (World Scientific, 1998)}
}

\author{T. JACOBSON }

\address{Department of Physics, University of Maryland\\
College Park, MD 20742-4111 USA}


\maketitle\abstracts{
A brief survey of the major themes and developments of 
black hole thermodynamics in the 1990's is given, 
followed by summaries of the talks on this subject at MG8
together with a bit of commentary, and closing with 
a look towards the future.
}
  
\section{Black hole thermodynamics in the 1990's}
The subject of black hole thermodynamics was born some twenty-five
years ago and it is still a source of much hope and mystification.
Except for a lull during the 1980's, there has been vigorous research
activity pursuing the promise held out by this confluence of gravitation,
quantum field theory, and thermodynamics. After initial incredulity
at Bekenstein's suggestion that the area of an event horizon really
is a measure of the entropy of the black hole, we have been led  
by a myriad of interconnecting results to the firm conclusion that
it is indeed so. 

It is the beauty of thermodynamics that no account
of the underlying microscopic details is required 
in order to deduce fundamental relations
between macroscopic quantities for
systems in (quasi-)equilibrium. This is why we feel we now have a 
foot in the quantum gravity door. However, to get another foot inside,
we need to pursue the microscopic description. This pursuit seems to 
amount to basically two questions: how does the gravitational
field microscopically react to the quantum process of Hawking radiation,
and what are the states which are counted by the black hole entropy? 

The resurgence of interest in black hole thermodynamics (BHT) 
in the 1990's
owes a lot to two waves of influence from string theory. 
The first wave came with the two-dimensional string-inspired
models of ``dilaton gravity". These models, coupled to matter,
are two-dimensional quantum field theories which possess black hole 
solutions that exhibit Hawking radiation
and which hold out the promise of being fully understandable 
if not exactly soluble.\cite{dilaton}
The second wave came with the introduction
of D-branes as nonperturbative stringy objects on which open
strings can terminate. D-branes made it possible to interpolate between
special highly supersymmetric black holes and perturbative
string physics. This correspondence has provided an account
for the black hole entropy on the stringy side in terms of an
enumeration of string states, and has provided 
a description of black hole
radiation in terms of unitary emission of closed strings 
from D-branes when pairs of open strings annihilate.\cite{dbranes}
The string results are in quantitatively precise agreement
with computations of entropy and Hawking flux based on the
utterly different framework of classical gravity and curved
space quantum field theory. This astonishing agreement 
may be simply a consequence of very powerful constraints
imposed by the high degree of supersymmetry, however
some detailed correspondence has also been found to extend
beyond the supersymmetric configurations.\cite{correspondence} 

Besides the string infusion, several other developments
have stimulated much research on BHT in the 1990's.
I hope the reader will forgive me for just mentioning
a few that have been particularly influential.
One is the extension
of black hole thermodynamics to allow for the higher curvature
couplings that arise in any low energy effective action (and in 
particular in the stringy one), both at the classical and quantum
levels.\cite{higherc} 
A related development was the important realization
that the divergences in the entropy of the thermal bath of 
acceleration radiation (or, (sometimes) equivalently, 
in the entanglement entropy of quantum field fluctuations 
across the event horizon) are precisely absorbed in the renormalization 
of Newton's
constant and the other parameters in the effective action that appear
in the expression for the black hole entropy.\cite{renormalization} 
This insight 
supports the recurring suspicion 
that black hole entropy is associated 
with vacuum fluctuations of quantum fields.

Another development of the 90's was sparked by the discovery
that black holes exist in three dimensional gravity with
a negative cosmological constant, 
a theory with no local degrees
of freedom at all which had already been extensively studied
as a model for quantum gravity.\cite{3dbh} The paradox of how 
these ``topological black holes" can have a huge entropy
seems sure to lead to important insights about quantum
gravity even in four dimensions. A partial  
understanding
of the state counting of ``would-be" gauge degrees of freedom
associated with the horizon has been achieved already.\cite{wouldbe}

More recently, an approach to black hole entropy is emerging
from the other major program in quantum gravity, loop
quantization. The area operator is perhaps the simplest and most
natural operator in this approach and is diagonal on the spin 
network states.
The area of a surface is determined by the spins on the lines
that puncture the surface, and the number of ways to obtain the
same area is the exponential of the area times an unknown constant.
The famous factor $1/4\hbar G$ is not yet accounted for in this approach,
since doing so would require somehow connecting the microscopic 
quantum gravity theory to the low energy effective Newton constant, 
which is something that has not yet been achieved. 
To make this connection would appear to require
dealing with the dynamics of the theory, an aspect
that is not yet understood. Perhaps the black hole
problem will be of some help in making that connection.

It is curious how utterly different the accounts of black
hole entropy seem to be according to the different lines
of approach being taken.  
This may be simply because one is only testing self-consistency
of theories, and nothing is being learned about Nature. 
However, it could be that one of these points of view 
will eventually make itself known as the right one,
perhaps the only one that really works, thus guiding
us towards the right theory of quantum gravity. More
likely, however, seems the possibility that the
commingling of these ideas will lead to a synthesis that
transcends any one of them.

\section{Black hole thermodynamics at MG8}

Four plenary talks were devoted to black hole thermodynamics  
at MG8. Bekenstein spoke on the  
the hypothesis that the black hole mass has a discrete spectrum yielding
a uniformly spaced area spectrum, which leads to a proportionality between
area and entropy and to a distortion of the semiclassical Hawking spectrum.
Parentani argued that when the gravitational field
is allowed to react, transition amplitudes for {\it matter}
are given by differences of the total {\it gravitational} action
in the WKB approximation. He illustrated this using both  
(homogeneous) quantum cosmology and Schwinger pair
creation. In the latter case, the gravitational action change
is one-fourth the change in area of the acceleration
horizon, which demonstrates that not just black hole horizons
but also acceleration horizons possess a thermodynamic
entropy which plays a role in dynamics, something which 
was first noticed in the context of black hole pair creation.
Maldacena spoke on black holes and D-branes, 
and Teitelboim spoke on 2+1 dimensional black holes.

With a couple of exceptions, the talks at the BHT parallel
sessions were not closely related to the subjects
of any of these four plenary talks, although most
were related somehow to one or more of the themes  
sketched above. This serves to emphasize that there
are quite a few active lines of approach to BHT.

Fifteen talks were given at the BHT sessions, which spanned
a total time of about six hours. All attendees who submitted 
abstracts spoke. 
(Abstracts were also submitted by five people who ultimately
were unable to attend the meeting.)
I will briefly describe
these talks here, grouping them into the two categories
of {\it Entropy} and {\it Hawking radiation and back-reaction}.
Where the titles of the contributed papers are different from 
those of the corresponding talks at MG8, I use here the paper titles
so as to facilitate cross referencing in these proceedings. 
In the case of multiple authors, the one who spoke is indicated
with an asterisk. References are given below only where there
is no associated contribution to these proceedings.

\subsection{Entropy}

Six of the talks dealt with microscopic accounts of 
the black hole entropy, while two focused on  
macroscopic aspects of the concept of black hole entropy 
extended beyond the usual black hole setting. 
In the order discussed here these were:
\begin{verse}
{} Induced entropy of a black hole in Sakharov's induced gravity\\
{\it V.P. Frolov}
 
{}  Quantum entropy of charged rotating black holes\\
{\it R.B. Mann*} and {\it S. Solodukhin} 
  
{} Black hole entropy and entanglement thermodynamics\\
{\it H. Kodama*, S. Mukohyama}, and {\it M. Seriu}

{} Rindler space entropy,
{\it J.R.A. Salazar*} and {\it J.M.T. Sarmiento}

{} Black hole entropy from loop quantum gravity,
{\it C. Rovelli}

{} The black hole entropy: a spacetime foam approach,
{\it F. Scardigli}
 
{} Black holes of constant curvature, {\it M. Banados}
    
{} The entropy of instantons with NUT charge\\
{\it S.W. Hawking and C.J. Hunter*}
    
\end{verse}
 
The first four talks listed above were concerned with the
idea of attributing the black hole entropy to the 
entropy of quantum fields associated with horizons.
This entropy, which goes by various names, can be viewed
either as the entanglement entropy across the horizon,
or as the entropy of the thermal bath of acceleration
radiation (i.e. Rindler or Boulware quanta) outside the
horizon. For certain types of matter fields (minimally
coupled scalars and spinor fields) this definition of the
matter field entropy is adequate. However, for nonminimally
coupled scalars or vector fields, one has extra contributions 
to the entropy of the gravitating canonical ensemble which arise
from the variation of the free energy with respect to the
temperature dependence of the mass of the background 
black hole.\cite{nonmin} It seems to me quite natural to include
this variation in the definition of the entropy of the
acceleration radiation, though it is not so easy to see
how to modify the notion of entanglement entropy---which 
is an information-theoretic entropy---to accommodate the 
extra contributions for nonminimally coupled scalars and vectors.
One idea is to presume that fields of this type are composites,
which resolve at very short distances into fundamental
scalars or spinors.\cite{composites} Another idea is that if the notion
of entanglement is somehow extended to take into account the 
quantum fluctuations of the horizon, then the extra
contributions to the entropy might be incorporated into 
the entanglement entropy. It should be emphasized, however,
that the contribution to the entropy in question can even be negative, so 
the generalization of the notion of entanglement entropy 
could not be just the information-theoretic entropy
of a single density matrix.

Frolov discussed a model ``induced gravity" field theory
in which the bare inverse Newton constant $G_B^{-1}$ vanishes and
the divergent contributions to the renormalized coupling $G_R^{-1}$
cancel at one loop so that $G_R^{-1}$ is finite.\cite{frolov} (To simplify
matters the higher derivative terms were ignored in this 
treatment but they could be handled similarly.) In this theory
the black hole entropy arises entirely from quantum matter field
fluctuations, so an interpretation of the entropy in terms of
a counting of quantum field states should be possible.
A straightforward interpretation in terms of the
entropy $S^{SM}$ of the acceleration radiation outside the black hole 
is frustrated by the contributions
to the entropy from the nonminimal couplings $\xi\phi^2 R$ 
(cf. the discussion in the previous paragraph),
which are required to obtain a finite result for $G_R^{-1}$.
Frolov argued that the discrepancy occurs because $S^{SM}$ 
counts the number of states
at a fixed value of the {\it hamitonian} whereas the black hole
entropy counts the number of states with a given {\it energy}.
The hamiltonian differs from the energy in this theory by a boundary
term proportional to the non-minimal couplings $\xi$, which
accounts for the difference $S^{BH}-S^{SM}$.  
I might emphasize that Frolov's approach here is to compute
a {\it microcanonical} entropy via a state counting. If one instead
computes the entropy of the {\it canonical} ensemble, I believe---as
Frolov himself first argued---that the extra terms arising from the
non-minimal couplings are produced as a result of the variation
of the black hole mass as a function of the temperature of the 
ensemble. The relation between the microcanonical and canonical 
viewpoints in this context deserves to be clarified.

Mann spoke on work which extends to the case of rotating black holes
prior results which established that, for static black holes,
the entropy of acceleration radiation of minimally coupled
massless scalar fields is precisely accounted for by the one loop 
renormalization of the gravitational action, including the
divergent parts. An interesting detail he mentioned is that
there are divergences that cannot be absorbed into curvature
counter-terms in the action but which ``cancel non-trivially"
on the Kerr-Newman backgrounds. I would
guess this cancellation can be understood as a general consequence
of the fact that the background field equations are satisfied.
 
Kodama's talk explored the proposal that the black
hole entropy is nothing but the entanglement entropy
of vacuum fluctuations across the horizon. Introducing
also an ``entanglement energy", which (according to one
of two definitions considered) is the difference of 
the energy in the original state and in the reduced state with
the correlations across the horizon (or other dividing surface)
taken out, he defined
an entanglement temperature $T_{\rm ent}=dS_{\rm ent}/dE_{\rm ent}$.
The calculations are done with a short distance cutoff $a$ in place.
This temperature $T_{\rm ent}$ 
diverges as $1/a$ in flat spacetime. However,
in a black hole spacetime, if the entanglement entropy is
first redshifted from to infinity from a proper distance $a$
from the horizon, then the temperature comes out to be of order
the Hawking temperature.
This result is closely 
related to the frameworks discussed by Frolov and Mann.
It seems particularly close in spirit to Frolov's identification
of the entropy as arising from counting 
the states with the same quantum field energy outside the horizon,
although the non-minimal coupling plays no role in Kodama's work.
This is a good place to mention also Salazar's talk, in which
the method of obtaining a reduced density matrix for the
quantum fields in Rindler spacetime was discussed. 

Rovelli presented a computation of the black hole entropy
in loop quantum gravity along the lines mentioned in section 1
above.\cite{rovelli}
The computation itself is straightforward, involving
the number of ways to obtain a given area with spin networks
puncturing a surface. Most of Rovelli's talk concerned rather
the conceptual underpinning of this calculation, addressing the
question why should the area be held fixed in the state counting.
He argued that only the shape of the horizon matters since the
inside is unobservable, and that fixing the area 
corresponds to fixing the energy, which defines the
microcanonical ensemble.  

Scardigli postulated a model for the states of a black hole 
which assigns the entropy to the degeneracy of configurations
of two-state ``topological cells" of Planck area which form the surface
of the horizon. He supposed that each cell can be found
in two different states, in one of which it carries no energy
and in the other its energy is the Planck mass. Using these ideas
he computed the entropy at fixed temperature for this system,
which for large black holes comes out proportional to the horizon
area.

Two of the talks focused on  
macroscopic aspects of the concept of black hole entropy 
extended beyond the usual black hole setting.  
Ba\~nados presented a family of constant curvature black holes
obtained by quotients of anti-de Sitter space in any dimension.
This construction generalizes that of the 2+1 dimensional
black hole discussed in Teitelboim's plenary talk. 
In the five dimensional case the energy and angular
momentum of these black holes can be defined as charges
of a Chern-Simons supergravity theory. Variations of these
charges satisfy a first law with an entropy that is {\it not}
proportional to the horizon area. I did not catch any
explanation of this surprising feature in the talk. 

A departure from the entropy-area relation also occurred
in Hunter's talk. He 
spoke on the idea of attributing entropy to 
stationary gravitational fields which are not black holes
but which do have zeroes of the Killing field,  
in particular four dimensional solutions to the Einstein
equations with NUT charge or magnetic mass. Although these
solutions can be assigned an entropy it is {\it not} related
to the (vanishing) area of the fixed point set.
 
\subsection{Hawking radiation and back-reaction}

Seven of the talks dealt with aspects of Hawking radiation
and the back reaction. In the order discussed here these
were:
\begin{verse}
{} Euclidean instantons and Hawking radiation, {\it S. Massar*} and 
{\it R. Parentani} 

{} Covariant path integrals and black holes, {\it F. Vendrell*}  
and {\it M.E. Ortiz} 
 
{} Loop corrections for 2D Hawking radiation, {\it A. Mikovi\'c*}  
and {\it V. Radovanovi\'c}
 
{}  The `ups' and `downs' of a spinning black hole\\
{\it C.M. Chambers*, W.A. Hiscock}, and {\it B.E. Taylor} 
                                         
{} Constraints on the geometries of black holes
in classical and semiclassical gravity,
{\it P.R. Anderson*} and {\it C.D. Mull} 
 
{} Semiclassical decay of near-extremal black holes, {\it T. Jacobson}

{} Thermodynamics of nonsingular spherically symmetric black hole\\
{\it I. Dymnikowa} 
\end{verse}

Massar described a quantum gravity
calculation in which the
rate of Hawking radiation
of charged shells by a charged black hole
is obtained from the difference of WKB actions 
of corresponding Euclidean instantons with and without the
charged shell, including the gravitational
back-reaction. This difference is equal to one fourth the 
difference in horizon areas, in agreement with 
Hawking's result.  
Only the {\it difference} of actions enters the rate,  
because one is  computing a dynamical transition amplitude,
so no ``regularization" of infinite actions is required. 
Massar argued that this relation between pair creation
rate and horizon area change---which was also discussed in
Parentani's plenary talk---applies quite generally
to all sorts of pair creation processes.  

Vendrell described a new mathematical approach 
to obtaining the thermal propagator for a particle in a black hole
spacetime. In this approach, only the exterior portion of the
spacetime is included in the configuration space, however
the tortoise coordinate is analytically continued to 
complex values. This yields a multiply connected complex
configuration space which covers the Kruskal manifold an
infinite number of times. The path integral for a particle
on this space yields the propagator in the Unruh vacuum.

Chambers presented results of numerical calculations which
revealed a counterexample to the usual expectation that a
rotating black hole will spin down to a final asymptotic
nonrotating state. In particular, a lone massless scalar field
radiates enough power in the $l=0$ mode so that the 
asymptotic value of $J/M^2$ is approximately 0.555 rather
than zero, whether it begins above or below this value. 
(In Nature, however, the presence of higher spin fields
would spin down the black hole completely unless there
is a very large number of hitherto undiscovered massless scalars.)
I might add that, curiously,  
this result seems to suggest that a nonrotating black hole
coupled only to massless scalars
is unstable to spinning up to just this value of $J/M^2$.

The subject of Mikovi\'c's talk was two dimensional black hole
evaporation in the CGHS dilaton gravity model. By adopting 
the point of view of reduced phase space operator quantization,
he is trying to take the study of this model further than has 
been previously possible.
He presented a loop expansion of the metric expectation value,
and argued that at two loops one can see, both from the
Bogoliubov coefficients in the effective geometry 
and from the (more reliable) flux operator, a non-classical 
increase in the Hawking temperature at late times. At still
later times the operator flux is  non-thermal,
drops to negative values,  and then approaches zero.
He sees indications that the higher loop corrections can remove the
singularities associated with the lower order approximations.

Anderson reported on constraints on 
static spherically symmetric black holes imposed
by the assumption that the curvature components
in the static orthonormal frame are analytic functions
of $r$ at the horizon. For conformally coupled free massless
fields the additional constraint imposed by the
trace of the semiclassical back-reaction equations implies 
that extremal black holes do not exist for a certain range 
of horizon radii. The excluded range depends on the number and
types of quantum fields, and can extend all the way to zero,
in which case there is a minimum allowed radius for an extremal 
black hole.

Jacobson's talk concerned the semiclassical description
of decay of a near-extremal black hole down to an extremal
state in the adiabatic approximation.\cite{jacobson} The motivation
was to try to reconcile the fact that near-extremal 
D-branes illuminated
by a pure state energy flux do not radiate unlimited entropy 
whereas the corresponding black holes seem to do so.
He argued that the semiclassical physics is very different
than for non-extremal black holes, due to a ``pile-up" of
the partners of Hawking radiation inside the horizon. However,
this difference does not appear sufficient to invalidate
the semiclassical analysis of the radiated entropy. Thus
no reconciliation with the D-brane analysis was achieved.

Dymnikowa's talk addressed the idea that the back-reaction
near a black hole singularity might remove the singularity,
replacing it by a de Sitter phase inside. 
She postulated a simple form for the stress-energy tensor
in regions of large curvature, motivated by an analogy with 
vacuum polarization in an electric field,
that would achieve this sort of configuration if matched
onto a Schwarzschild solution. Using this form she argued that
an evaporating black hole would shrink down to a critical 
nonvanishing size at which its temperature would vanish.
To determine the further evolution of this ``extremal state" 
would seem to require a more complete dynamical picture. 
 
\section{Black hole thermodynamics in the future}

Despite the advances of recent years, much remains to be 
understood about black hole thermodynamics. 
In the long run, the goal is nothing short of a full understanding
of quantum gravity and what it means for singularities, the fate of
black holes, unitarity, and the topology of spacetime. The day when 
this goal is met seems very far off indeed. However,
there is plenty of reason for optimism in the short run. 
We can anticipate improved understanding of
the state counting entropy of 2+1 dimensional and other
topological black holes, including the 1+1 dimensional dilatonic 
topological case, which has so far resisted attempts.\cite{1+1}
A much
better understanding of 1+1 dilaton gravity coupled to matter
seems possible, including a definitive statement about the
fate of singularities and unitarity. 
If string theory can stay
put long enough it should be possible for someone to understand 
why semiclassical black hole physics permits unlimited (entanglement) 
entropy production whereas D-brane physics does not. String
theory (and its descendents) will continue to provide new insights,
perhaps via the matrix theory description of neutral black holes
and Hawking radiation.\cite{matrix} The loop quantization approach may well 
be successfully extended to count the entropy of rotating black holes,
and perhaps the factor of 1/4 will be understood together with something
of the dynamics in this formulation. Condensed matter analogies for
black holes may provide observable instances of the Hawking effect
and the decay of the ergoregion\cite{condmat}, and it should be possible 
to understand in detail how the outgoing black hole modes are
produced without an infinite density of states at the horizon in these
systems\cite{lattice}.
It seems not unreasonable to expect that many of these and other puzzles
will be resolved in the next several years, some of them in time for the
next Marcel Grossmann meeting.

\section*{Acknowledgments}
I would like to thank the organizers of MG8 and the
speakers in the black hole thermodynamics parallel sessions for 
their contributions.  
The participation and work of this author were supported 
by NSF travel grant PHY97-22529 (to North Carolina State 
University) and NSF grant PHY94-13253.

\section*{References}
References are given to review articles whenever possible rather than 
to original sources.

\end{document}